\documentclass[conference]{IEEEtran}
\IEEEoverridecommandlockouts
\usepackage{cite}
\usepackage{amsmath,amssymb,amsfonts}
\usepackage{algorithmic}
\usepackage{graphicx}
\usepackage{textcomp}
\usepackage{xcolor}
\def\BibTeX{{\rm B\kern-.05em{\sc i\kern-.025em b}\kern-.08em
    T\kern-.1667em\lower.7ex\hbox{E}\kern-.125emX}}

\usepackage{cite,hyperref}
\usepackage{amsmath,amssymb,amsfonts}
\usepackage{algorithm}
\usepackage{algorithmic}
\usepackage{multirow}
\usepackage{subfigure}  
\usepackage{makecell}
\usepackage{booktabs}

\usepackage{multirow}

\usepackage{graphicx}
\usepackage{textcomp}
\def\BibTeX{{\rm B\kern-.05em{\sc i\kern-.025em b}\kern-.08em
    T\kern-.1667em\lower.7ex\hbox{E}\kern-.125emX}}
\begin{document}

\title{Defensive Distillation based Adversarial Attacks Mitigation Method for Channel Estimation using Deep Learning  Models in Next-Generation Wireless Networks}


\author{\IEEEauthorblockN{Ferhat Ozgur Catak}
\IEEEauthorblockA{\textit{Department
of Electrical Engineering \& Computer Science} \\
\textit{ University of Stavanger}\\
Rogaland, Norway \\
f.ozgur.catak@uis.no}
\and
\IEEEauthorblockN{Murat Kuzlu}
\IEEEauthorblockA{\textit{Department
of Engineering Technology} \\
\textit{Old Dominion University,}\\
Old Dominion University, Norfolk,
VA, USA \\
mkuzlu@odu.edu}
\and
\IEEEauthorblockN{Evren Catak}
\IEEEauthorblockA{\textit{Independent researcher}\\
Stavanger, Norway  \\
evren.catak@ieee.org}
\and
\IEEEauthorblockN{Umit Cali}
\IEEEauthorblockA{\textit{Department of Electric Power Engineering,)} \\
\textit{Norwegian University of Science and Technology}\\
Trondheim, Norway \\
umit.cali@ntnu.no}
\and
\IEEEauthorblockN{Ozgur Guler}
\IEEEauthorblockA{\textit{eKare Inc.}\\
Fairfax, VA, USA, \\
oguler@ekare.ai}
}




\maketitle

\begin{abstract}
Future wireless networks \textcolor{black}{(5G and beyond),} also known as Next Generation or NextG, are the vision of forthcoming cellular systems, connecting billions of devices and people together. In the last decades, cellular networks have been dramatically growth with advanced telecommunication technologies for high-speed data transmission, high cell capacity, and low latency. The main goal of those technologies is to support a wide range of new applications, such as virtual reality, metaverse, telehealth, online education, autonomous and flying vehicles, smart cities, smart grids, advanced manufacturing, and many more. The key motivation of NextG networks is to meet the high demand for those applications by improving and optimizing network functions. Artificial Intelligence (AI) has a high potential to achieve these requirements by being integrated in applications throughout all layers of the network. However, the security concerns on network functions of NextG using AI-based models, i.e., model poising, have not been investigated deeply. It is crucial to protect the next-generation cellular networks against cyber attacks, especially adversarial attacks. Therefore, it needs to design efficient mitigation techniques and secure solutions for NextG networks using AI-based methods. This paper proposes a comprehensive vulnerability analysis of  deep learning (DL)-based channel estimation models trained with the dataset obtained from MATLAB’s 5G toolbox for adversarial attacks and defensive distillation-based mitigation methods. The adversarial attacks produce faulty results by manipulating trained DL-based models for channel estimation in NextG networks, while making models more robust against any attacks through mitigation methods. This paper also presents the performance of the proposed defensive distillation mitigation method for each adversarial attack against the channel estimation model. The results indicated that the proposed mitigation method can defend the DL-based channel estimation models against adversarial attacks in NextG networks.
\end{abstract}


\section{Introduction}

\subsection{Preamble}  
In the last decade, the next-generation networks deployed on cellular networks \textcolor{black}{(i.e., 5G and beyond) are undergoing a major revolution along with advanced telecommunication technologies for high-speed data transmission, high cell capacity, and low latency. Each network has own focus, i.e., 5G: deliver higher multi-Gbps peak data speeds, ultra low latency, 6G: embed artificial intelligence.} NextG networks require a high cost investment and research to meet infrastructure, computing, security
and privacy requirements. These technologies will enable the next data communications and networking era by connecting everyone to a world in which everything, everywhere is connected.  The main goal of those technologies is to support a wide range of new applications, such as Augmented reality (AR), Virtual reality (VR), metaverse, telehealth, education, autonomous and flying vehicles, smart cities, smart grids, and advanced manufacturing. They will create new opportunities for industries to improve visibility, enhance operational efficiency, and accelerate automation \cite{bertino2021computing}. It is expected that next-generation networks must simultaneously provide high data speed, ultra low latency, and high reliability to support services for those applications \cite{7247338}.  Artificial Intelligence (AI) plays a crucial role to achieve these requirements by being integrated in applications throughout all levels of the network. AI is one of key drivers for next-generations wireless networks to improve the efficiency, latency, and reliability of the network applications \cite{8539642}. AI is also applied into the channel estimation applications, which is one of the fundamental prerequisites in  wireless networks. The traditional channel estimation methods are extremely complex and low accurate due to the multi-dimensional data structure and the nonlinear characteristics of the channel. Therefore, DL-based channel estimation models have been used in next-generation networks to address the traditional channel estimation. However, DL-based channel estimation models can be vulnerable to adversarial machine learning (ML) attacks. Therefore, a secure scheme is crucial for DL-based channel estimation models used in next-generation networks, in addition to the security and vulnerability issues. DL-based models in the next-generation wireless communication systems should be evaluated before deploying them to the production environments in terms of vulnerabilities, risk assessment, and security threats.

\subsection{Related Works}
The main goal of NextG networks is to provide very high data rates (Tbps) and extremely low latency (less than milliseconds) with a high cell capacity (10 million devices for every square kilometer) \cite{7414384, 9083885}. The key of next-generation network is to use new technologies, such as millimeter wave (mmWave), massive multiple-input multiple-output (massive MIMO), and AI. mmWave is essential for those networks, which provides  a high capacity, throughput and very-low latency  in frequency bands above 24 GHz. Massive MIMO is an advanced version of MIMO, which inludes a group of antennas at both the transmitter and receiver side. This method provide a better throughput and spectrum efficiency in wireless communication. AI-based algorithm have been used to improve network performance and efficiency. This study focuses on DL-based channel estimation models in
next-generation wireless networks, and their vulnerabilities.  In the literature, these topics have already studied with and without vulnerability concerns \cite{8808168, kaur2021machine, wilhelmi2021usage, yazar2020waveform, khan2022efficient, piran2019learning}. The authors in \cite{8808168} reviewed  AI-empowered wireless networks and the role of AI in deploying and optimizing next generation architectures in terms of operations. It indicated that AI-based models have already used to train the transmitter, receiver, and channel  as an auto-encoder. This allows the transmitter and receiver to be optimized mutually. The study also indicated that next-generation networks will be different from current ones in many ways, such as network infrastructures, wireless access technologies, computing, application types, etc. The authors of \cite{ozpoyraz2022deep} reviewed DL-based solutions in next generation networks, focusing on physical layer applications of cellular networks from massive MIMO, reconfigurable intelligent surface (RIS), and multi-carrier (MC) waveform. It also emphasized the AI-based solutions' contribution to the improvement of the network performance. The authors in \cite{8815888, 8896030} proposed a robust channel estimation framework using the fast and flexible denoising convolutional neural network (FFDNet) and deep convolutional neural networks (CNNs) for mmWave MIMO. Both proposed methods can deal with a wide range of signal-to-noise ratio (SNR) levels with a flexible noise level map, and offer a better performance for channel estimators in terms of the accuracy. DL-based algorithms have a significant contribution to improving the overall system performance for next-generation wireless networks. Fortunately, the main potential security issue related to AI-based algorithms, i.e., model poising, is studied by several research groups in the wireless research community \cite{dang2020should, kuzlu2021role}. The authors of \cite{porambage20216g, siriwardhana2021ai} provided a comprehensive review of NextG wireless networks in terms of opportunities, and security and privacy challenges as well as proposes solutions for NextG networks. 

\subsection{Purpose and Contributions}
The channel estimation and the interference are extremely convoluted topics in the wireless communication due to dynamic nature of its communication channel. AI-based models can automatically extract the channel information by learning channel characteristics from previous communication channel data. However, it is difficult to find the correlation between many resources, system parameters, and dynamic communication environment by using existing techniques. Therefore, sophisticated AI-based algorithms can help to model the highly nonlinear correlations and estimate the channel characteristics \cite{wang2020artificial}.  In our recent papers \cite{catak2022security} and \cite{9527756}, adversarial attacks and mitigation methods have been investigated along with the proposed framework for mmWave beamforming prediction models in next-generation networks. This paper implements widely used adversarial attacks from Fast Gradient Sign Method (FGSM), Basic Iterative Method (BIM), Projected Gradient Descent (PGD), Momentum
Iterative Method (MIM), to Carlini \& Wagner (C\&W) as well as a defensive distillation based mitigation method for DL-based channel estimation models in next-generation wireless networks. The results showed that DL-based models used in these networks are vulnerable to adversarial attacks, while the models can be more secure against adversarial attacks through the proposed mitigation method. \textcolor{black}{The code is available from GitHub} \footnote{\url{https://github.com/ocatak/6g-channel-estimation-dataset}}.

\section{Preliminaries} \label{sec:preliminaries}
In this section, a brief overview of the channel estimation, the adversarial ML attacks, such as FGSM, BIM, PGD, MIM, and C\&W along with defensive distillation based mitigation, are presented. Dataset description and scenarios are also given with the selected performance metrics to evaluate the models' performance under normal and attack conditions.

\subsection{Channel Estimation for Communication System}
In wireless communication system, the channel characteristic presents the communication link properties between transmitter and receiver. It is also known as channel state information (CSI). The signal is transmitted through a communication channel. i.e., medium, and the transmitted signal is received as a distorted and noise added. It is needed to decode the received signal and remove the unwanted signal, i.e., distortion and noise added by the channel, from the received signal. To identify the channel characteristics is the  first process to achieve that, which is called \textit{channel estimation} process. The received signal is attenuated by a factor \(h_0\) and delayed by a specific time \(\tau_0\). \(h_0\) depends on the propagation medium, frequency, $Tx/Rx$ gains, while \(\tau_0\) depends on the speed of electromagnetic wave in the medium. 

It is assumed that $x(t)$ presents the transmitted signal, while \(y(t)\) presenting the received signal. When \(x(t)\) is transmitted through communication channel, i.e., air, the signal is distorted, and noise is added with the transmitted signal. As a result, the received signal y(t) is not the same as the transmitted signal \(x(t)\). Received signal \(y(t)\) is shown as:

\begin{equation}
    y(t) = h_0 * x(t - \tau_0)
\end{equation}

However, the received signal is composed of several reflected and scattered paths, i.e., multiple paths, with a different attenuation and delay. The composed received signal is shown as:

\begin{equation}
    y(t)=\sum_{l = 0}^{l}h_l * x(t - \tau_l)
\end{equation}
where \(l\) is the specific path/tap at a time. 

The mobility causes Doppler frequency shift, i.e., the change in the wavelength or frequency of the waves as to the observer being in motion with respect to the wave source.  Doppler effect plays a important role in telecommunications and computations of signal path loss and fading due to multi-path propagation. In addition, the channel characteristics, i.e., \(h_0\) an \(\tau_0\), can also change over time due to the mobility of the one of communication sides, and shown as {\(h_l^t\)} and {\(\tau_l^t\)}.  The channel can characterized by number of paths/taps and dependence of channel coefficients and delay on time. The final received signal with the Doppler effect can be shown at a specific time as: 

\begin{equation}
    y(t)=\sum_{l = 0}^{l}h_l^t * x(t - \tau_l^t)
\end{equation}

The channel estimation plays an important part in a wireless communications for increasing the capacity and the overall system performance. It is a high demand for new wireless networks along with higher data rate, better quality of service, and higher network capacity. Therefore, it is needed new promising technologies to meet these requirements. A migration from Single Input Single Output (SISO) to Multiple Input Multiple Output (MIMO) antenna technology has started with NextG networks. \textcolor{black}{The channel estimation is the core of next-generation communication systems, 5G and beyond, performed in different ways for SISO and MIMO approaches at the receiver side. The channel estimation algorithm can be classified into three main categories, i.e., blind channel estimation, semi-blind channel estimation, and training-based estimation
\cite{oyerinde2012review}. The training-based estimation is widely used in the transmission.
The general approach of this estimation is to insert known reference symbols, i.e., pilots, into the transmitted signal and then interpolate the channel response based on these known pilot symbols.} The  process works as the following steps: 
(1) develop a mathematical model to correlate the transmitted and received signals  using channel characteristics, (2) embed a predefined signal, i.e., pilot signal, into the transmitted signal, (3) transmit the signal through the channel (4) receive transmitted signal as a distorted and/or noise added through the channel, (5) decode the pilot signal from the received signal, and (6) compare the transmitted and the received signals, and (7) find correlation between the transmitted and the received signals.

A channel model is a representation of the channel that a transmitted signal follows to the receiver. In the simulation environment, the channel model is typically classified into two categories, i.e, clustered delay line (CDL) model and tapped delay line (TDL) channel model. A CDL is used to model the channel when the received signal consists of multiple delayed clusters. Each cluster contains multipath components with the same delay, but slight variations for angles of departure and arrival, i.e., MIMO. On the other hand, a TDL model is  defined as a simplified evaluations of CDL, i.e., non-MIMO evaluations or SISO.  These channel models are defined well in technical report  released by 3GPP, i.e., 3rd Generation Partnership Project \cite{3GPP_v1}.  According to this report, CDL/TDL models are defined in the frequency range from 0.5 GHz to 100 GHz with a maximum bandwidth of 2 GHz. For CDL/TDL models, five different channel profiles  models are constructed, i.e.  A, B, and C for non-line-of-sight (NLOS) progation  while D and E for line-of-sight (LOS) propagation. Power, delay and angular information are used to define CDL models, while power, delay and Doppler spectrum information are used for TDL models in the technical report released by 3GPP.  

\textcolor{black}{
There have been many efforts regarding the channel estimation algorithms using in the literature. However, it is still a challenging problem due to the computational complexity degree of algorithms along with enormous amount of mathematical operations and low performance in terms of accuracy. The equalization method is typically used to reduce the complexity at
the receiver side \cite{saad2021enhanced}. With the introducing the machine learning methods to 5G and beyond applications, it is expected that those methods can improve the performance of the channel estimation algorithm in terms of the degree of low computational complexity and accuracy.  It is clear that the accuracy of the channel estimation can be improved with the deep learning method compared to conventional channel estimation algorithms \cite{yang2019deep}. In addition, the dynamic nature of deep learning algorithms can also save a considerable amount of computational power for complex analysis needed in channel estimation algorithm \cite{simeone2018very}. However, it can be questionable of the feasibility of using machine learning methods in channel estimation. The authors used a CNN combined with a projected gradient descent algorithm to demonstrates the feasibility of using machine learning methods in channel estimation. }

\subsection{Convolutional Neural Networks}
\label{sec:cnn}

The convolutional neural network (CNN) is a neural network that has shown to be very successful for image recognition \cite{fukushima1982neocognitron, lecun1989backpropagation, 5537907}.
Compared to the fully-connected neural network, CNN can extract all the information with a lower number of parameters.
The main idea of the CNN is that we can locate the structure of an image by the convolution operation.
Suppose that the image $\mathbf{x}$ is a two-dimensional matrix.
The convolution operation between the image $\mathbf{x}$ and a filter $\mathbf{W}$ is defined by
\begin{equation}
\label{eq:conv}
\mathbf{y} = \mathbf{W} \ast \mathbf{x} = \sum_{i=1}^{W} \sum_{j=1}^{H} \mathbf{W}_{i,j} \mathbf{x}_{i-s,j-s},
\end{equation}
where $W$ and $H$ are the width and height of the image $\mathbf{x}$, respectively, and $s$ is the number of stride, which is the distance between two adjacent positions.

The CNN is composed of several types of layers.
The convolution layer is the most critical layer of the CNN, consisting of several filters.
Each filter extracts a particular type of feature from an input image.
The pooling layer is a down-sampling layer, which is used to reduce the size of the convolution output.
Each pooling operation replaces several adjacent values with the maximal value or the mean value.
The fully-connected layer is a standard neural network layer used to combine all the features extracted by the convolution layer.
The softmax layer is a classification layer to classify the input data.

The input image is a two-dimensional matrix.
The filter in the convolution layer extracts a particular type of feature from the input image.
For example, the leftmost filter extracts horizontal lines, and the middle filter extracts diagonal lines.
The output of the convolution layer is then sent to the pooling layer, which reduces the size of the data.
The output of the pooling layer is then sent to the fully-connected layer, which combines all the features extracted by the convolution layer.
The output of the fully-connected layer is then sent to the softmax layer, which classifies the data.

\subsection{Adversarial Attacks}
ML-based models are trained to automatically learn the underlying patterns and correlations in data by using algorithms. Once an ML-based model is trained, it can be used to predict the patterns in new data. The accuracy of the trained model is essential to achieving a high performance, which can also be called as a generalization. However, the trained model can be manipulated by adding noise to the data, i.e., targeted and non-targeted adversarial ML attacks. The adversarial ML attacks are generated by adding a perturbation to a legitimate data point, i.e., an adversarial example, to fool the ML-based models. There are various kinds of adversarial ML attacks, such as evasion attacks, data poisoning attacks, and model inversion attacks \cite{lin2021ml}. 

An evasion attack aims to cause the ML-based models to misclassify the adversarial examples as legitimate data points, i.e., targeted and non-targeted evasion attacks. Targeted attacks aim to force the models to classify the adversarial example as a specific target class. Non-targeted attacks aim to push the models to classify the adversarial example as any class other than the ground truth. Data poisoning aims to generate malicious data points to train the ML-based models to find the desired output. It can be applied to the training data, which causes the ML-based models to produce the desired outcome. Model inversion aims to generate new data points close to the original data points to find the sensitive information of the specific data points.

Taking channel estimation CNN model as an example, here, we use $h(\mathbf{x},\omega): \mathbb{R}^{m \times n} \mapsto \mathbb{R}^{m \times n}$ to denote the channel estimation CNN model, where $\omega$ is the parameters of the channel estimation CNN model, and $\mathbf{x}$ is the input data. The objective of a targeted adversarial attack is to generate an adversarial example $\mathbf{x}'$ from a legitimate example $\mathbf{x}$ to fool the channel estimation CNN model to produce the desired output. The attacker uses the lowest possible budget to corrupt the inputs, aiming to increase the distance (i.e., MSE) between the model's prediction and the real channel. Therefore $\sigma$ is calculated as

\begin{equation}
    \sigma^* = \underset{|\sigma|_p \leq \epsilon}{\arg max}\,\,\ell(\omega,\mathbf{x}+\sigma,\mathbf{y})
\end{equation}
where $\mathbf{y} \in \mathbb{R}^{m \times n}$ is the label (i.e., channel information), and $p$ is the norm value and it can be $0, 1, 2, \infty$.

\begin{figure*}[!htbp]
    \centering
    \includegraphics[width=1.0\linewidth]{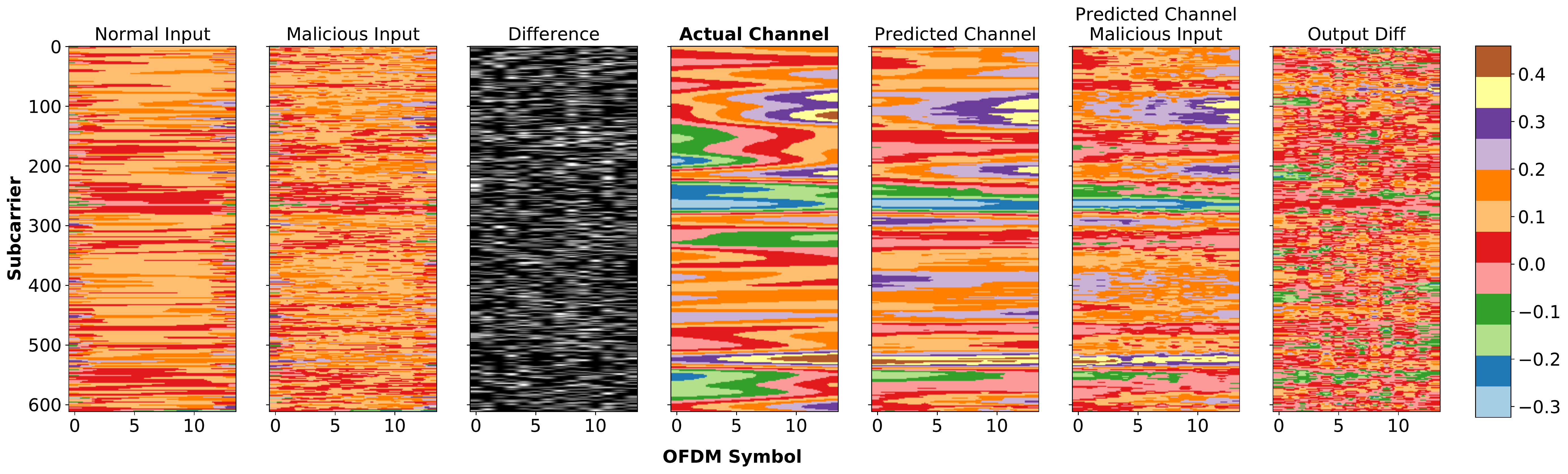}
    \caption{Typical adversarial ML-based adversarial sample generation}
    \label{fig:typical_attack2}
\end{figure*}

Figure \ref{fig:typical_attack2} shows a typical adversarial ML-based adversarial sample generation procedure.

These adversarial attack types are given as follows. 

\subsubsection{FGSM}
Fast Gradient Sign Method (FGSM): FGSM is one of the most popular and simplest approaches to construct adversarial examples. It is called one-step gradient-based attacks. It is used to compute the gradient of the loss function with respect to the input, $\mathbf{x}$ and then the attacker creates the adversarial example by adding the sign of the gradient to the input data \ It was first introduced by Goodfellow et al. in 2014 \cite{michels2019vulnerability}. 
The gradient sign is computed using backpropagation
algorithm. The steps are summarized as follows:

\begin{itemize}
    \item Compute the gradient of loss function, $\nabla_{\mathbf{x}}\ell(\mathbf{x},\mathbf{y})$
    \item Add the gradient to the input data, $\mathbf{x}_{adv} = \mathbf{x} + \epsilon \times sign(\nabla_{\mathbf{x}}\ell)$
\end{itemize}
where $\epsilon$ is the budget. FGSM attack has been used in \cite{2021arXiv210204150F} to attack channel estimation prediction models.

\subsubsection{BIM} 
Basic Iterative Method (BIM):  BIM is one of the most popular attacks, which is called iterative gradient-based attacks. This attack is derived from the FGSM attack. It is used to compute the gradient of the loss function with respect to the input, $\mathbf{x}$ and then the attacker creates the adversarial example by adding the sign of the gradient to the input data.  The gradient sign is computed using backpropagation algorithm. The steps are summarized as follows:

\begin{itemize}
    \item Initialize the adversarial example as $\mathbf{x}_{adv} = \mathbf{x}$
    \item Iterate $i$ times, where $i=0, 1, 2, 3,..., N$
    \item Compute the gradient of loss function, $\nabla_{\mathbf{x}}\ell(\mathbf{x}_{adv},\mathbf{y})$
    \item Add the gradient to the input data, $\mathbf{x}_{adv} = \mathbf{x}_{adv} + \epsilon \times sign(\nabla_{\mathbf{x}}\ell)$
\end{itemize}
where $\epsilon$ is the budget and $N$ is the number of iterations. The BIM attack has been used in \cite{2021arXiv210204150F} to attack channel estiomation prediction models.

\subsubsection{PGD} PGD is one of the most popular attacks, which is called gradient-based attacks \cite{jiang2021project}. It is called the most powerful attack. It is used to compute the gradient of the loss function with respect to the input, $\mathbf{x}$ and then the attacker creates the adversarial example by adding the sign of the gradient to the input data. The gradient sign is computed using backpropagation algorithm. The steps are summarized as follows:

\begin{itemize}
    \item Initialize the adversarial example as $\mathbf{x}_{adv} = \mathbf{x}$
    \item Iterate $i$ times, where $i=0, 1, 2, 3,..., N$
    \item Compute the gradient of loss function, $\nabla_{\mathbf{x}}\ell(\mathbf{x}_{adv},\mathbf{y})$
    \item Add random noise to the gradient, $\hat{\nabla}_{\mathbf{x}}\ell(\mathbf{x}_{adv},\mathbf{y}) = \nabla_{\mathbf{x}}\ell(\mathbf{x}_{adv},\mathbf{y}) + \mathcal{U}(\epsilon)$
    \item Add the gradient to the input data, $\mathbf{x}_{adv} = \mathbf{x}_{adv} + \alpha \times sign(\hat{\nabla}_{\mathbf{x}}\ell)$
\end{itemize}
where $\epsilon$ is the budget, $N$ is the number of iterations, and $\alpha$ is the step size. PGD can generate stronger attacks than FGSM and BIM. 

\subsubsection{MIM} 
Momentum Iterative Method (MIM): MIM is a variant of the BIM adversarial attack, introducing momentum term and integrating it into iterative attacks \cite{fostiropoulosrobust}. 
It is used to compute the gradient of the loss function with respect to the input, $\mathbf{x}$ and then the attacker creates the adversarial example by adding the sign of the gradient to the input data. The gradient sign is computed using backpropagation algorithm. The steps are summarized as follows:

\begin{itemize}
    \item Initialize the adversarial example $\mathbf{x}_{adv} = \mathbf{x}$ and the momentum, $\mu = 0$
    \item Iterate $i$ times, where $i=0, 1, 2, 3,..., N$
    \item Compute the gradient of loss function, $\nabla_{\mathbf{x}}\ell(\mathbf{x}_{adv},\mathbf{y})$
    \item Update the momentum, $\mu = \mu + \frac{\eta}{\epsilon} \times \nabla_{\mathbf{x}}\ell(\mathbf{x}_{adv},\mathbf{y})$
    \item Add random noise to the gradient, $\hat{\nabla}_{\mathbf{x}}\ell(\mathbf{x}_{adv},\mathbf{y}) = \nabla_{\mathbf{x}}\ell(\mathbf{x}_{adv},\mathbf{y}) + \mathcal{U}(\epsilon)$
    \item Add the gradient to the input data, $\mathbf{x}_{adv} = \mathbf{x}_{adv} + \alpha \times sign(\hat{\nabla}_{\mathbf{x}}\ell)$
\end{itemize}
where $\epsilon$ is the budget, $N$ is the number of iterations, $\eta$ is the momentum rate, and $\alpha$ is the step size.

\subsubsection{Carlini \& Wagner}
The Carlini \& Wagner (C\&W) attack is based on the idea of a zero-sum game. In a zero-sum game, the total amount of value in the game is fixed. The winner of the game gets all of the value, and the loser gets nothing.  The C\&W method is an iterative attack that constructs adversarial examples by approximately solving the minimization problem $min_d(x,x')$ such that $f(x') = t'$ for the attacker-chosen target $t'$, where $d(\cdot)$ is an appropriate distance metric. The optimization problem is shown at the following equation:

$$min_{x \in \mathcal{X}} \mathbb{E}_{y \in \mathcal{Y}} [f(x) - y]^2$$

where $x \in \mathcal{X}$ is a training example, $y \in \mathcal{Y}$ is the target output, and $f(x)$ is the function to be estimated. The optimization is solved for a set of points $x'$ that are close to the target $t'$, such that the function $f(x) - y$ is maximized for all $y$. This produces a set of adversarial examples $x'$ that are likely to fool the defender model.

The most important difference between C\&W and other adversarial ML attacks is that C\&W does not require an $\epsilon$ value for the optimization. That is, C\&W does not require that the attacker's goal be to find a set of points that are close to the target, but rather finds a set of points that are guaranteed to fool the defender. This makes C\&W a more powerful attack.

\section{Defensive Distillation}
\begin{figure*}[!htbp]
 \centering
	\includegraphics[width=0.95\linewidth]{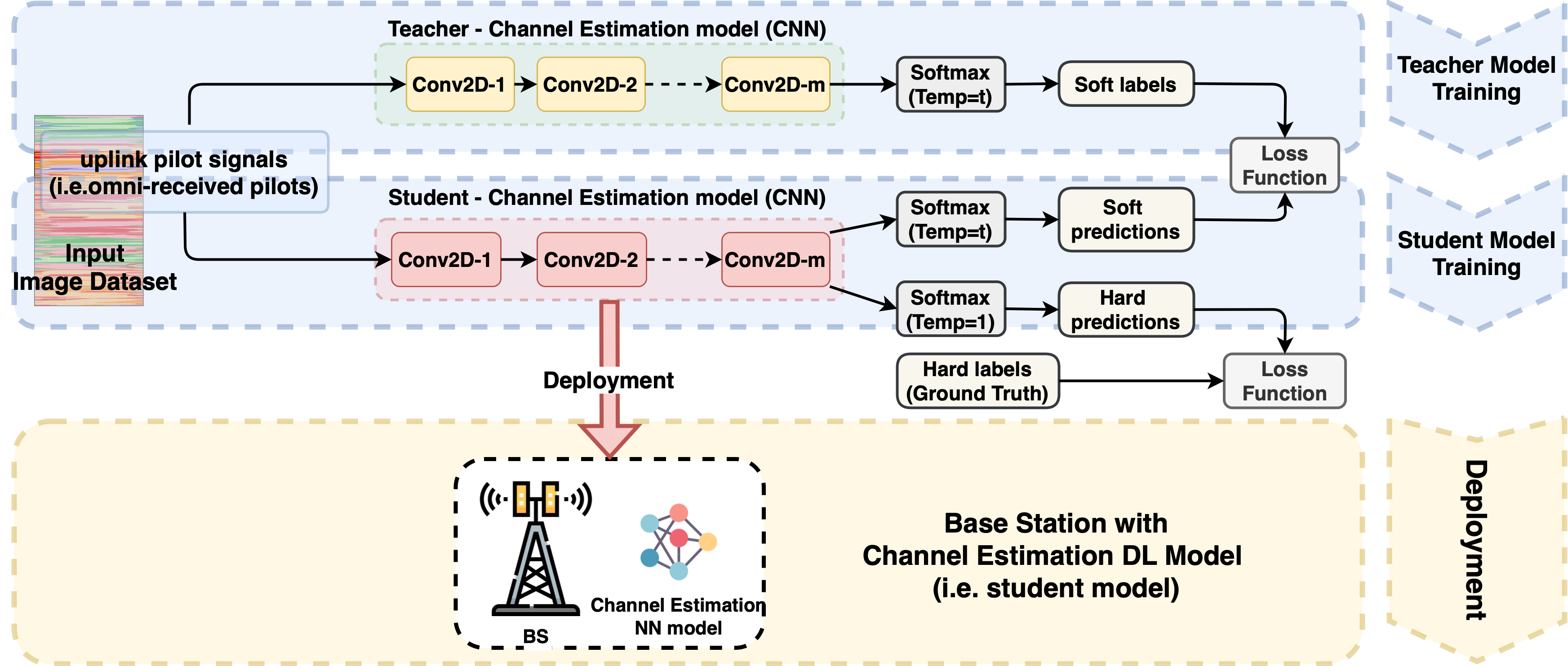}
	\caption{Overview of the system architecture with knowledge distillation. }
	\label{fig:knowledge_distillation_figure}
\end{figure*}

Knowledge distillation was previously introduced by Hinton et al. \cite{hinton2015distilling} to compress the knowledge of a large, densely connected neural network (the teacher) into a smaller, sparsely connected neural network (the student). It was shown that the student was able to reach a similar performance as the teacher \cite{hinton2015distilling}. In the initial work, the knowledge distillation was used to solve a classification problem, which is also called the teacher-student framework. Papernot et al. \cite{papernot2016distillation} proposed this technique for the adversarial ML defense and demonstrated that it could make the models more robust against adversarial examples. The main contribution of this work was to introduce the knowledge distillation to the adversarial ML defense. The defensive distillation is an ML framework that can enhance the robustness of the model for classification problems. The first step is to train the teacher model with a high temperature ($T$) parameter to soften the softmax probability outputs of the DL model. This can be done as follows:

\begin{equation}
    p_{softmax}(z, T) = \frac{e^{z/T}}{\sum_{i=1}^{n}e^{z_{(i)}/T}}
\end{equation}
where $n$ is the number of labels and $z$ is the output of the last layer of the DL model, i.e., $z = \mathbf{W}_n \cdot \mathbf{a}_{n-1} + b_n$. Here, $\mathbf{W}_n$ is the weight matrix and $\mathbf{a}_{n-1}$ is the activation of the last layer. In the second step, the softmax probability outputs are used to train the student model with a lower temperature parameter. The objective function is defined as

\begin{equation}
\begin{aligned}
    \mathcal{L}_{student}(T) &= \frac{1}{N} \sum_{i=1}^{N} \sum_{j=1}^{n} \mathbf{y}_{ij} \cdot \log p_{softmax}(z_{ij}, T) \\
    &= \frac{1}{N} \sum_{i=1}^{N} \sum_{j=1}^{n} \mathbf{y}_{ij} \cdot \log \frac{e^{z_{ij}/T}}{\sum_{i=1}^{n}e^{z_{ij}/T}}
\end{aligned}
\end{equation}
where $N$ is the number of training samples, $\mathbf{y}_{ij}$ is the training label, and $z_{ij}$ is the logit. The objective function for the training the teacher model is defined as

\begin{equation}
    \mathcal{L}_{teacher}(T) = -\frac{1}{N} \sum_{i=1}^{N} \sum_{j=1}^{n} \mathbf{y}_{ij} \cdot \log \frac{e^{z_{ij}/T}}{\sum_{i=1}^{n}e^{z_{ij}/T}}
\end{equation}
Defensive distillation is a method that can enhance the robustness of the models, which are trained by the soft targets provided by the teacher model. By minimizing the objective functions, the model can be trained. This method is a very useful tool for building robust models against adversarial examples \cite{papernot2016distillation}. Figure \ref{fig:knowledge_distillation_figure} shows the overall steps for this technique. According to the figure, the teacher model is typically a large and deep neural network, while the student model is usually a small and shallow neural network. The knowledge distillation process consists of two steps: (1) training the teacher model and (2) distilling the knowledge from the teacher to the student. The distillation can be performed using the teacher model's output probabilities, the teacher model's activations, or the intermediate representations of the teacher model. The distillation can also be performed using a distillation loss, typically a combination of the cross-entropy loss and the distillation loss. The cross-entropy loss is used to minimize the difference between the output probabilities of the teacher and student models. In contrast, the distillation loss is used to minimize the difference between the intermediate representations of the teacher and student models.

\textcolor{black}{In a typical communication system, the channel estimation is done by the base station with the help of pilot signals sent by the user equipment (UE) during uplink. And the base station sends pilot signals towards the UE, which acknowledges with the estimated channel information for the downlink transmission. Network operators and service providers are responsible for running their operation properly and meeting their obligations to the customers and the public related to privacy and data confidentiality. However, the network operations can be vulnerable to the machine learning adversarial attacks, especially 5G and beyond due to using machine learning-based applications. In the figure, the training of the channel estimation prediction model (i.e., student model) is protected against adversarial ML attacks and its use in base stations are shown in all its stages. The teacher model is trained as the first step, the student model is trained with the predictions made by the teacher model, and the real labels with the student model's predictions are used as the loss function inputs as the second step. In this way, the knowledge of the teacher model is compressed and transferred to the student model. The student model is deployed to the base stations in the last stage.}

Deep learning approaches have shown to perform extremely well for a wide range of computer vision tasks (e.g., image classification, object and action detection, scene segmentation, image generation, etc.). However, deep neural networks (DNNs) require large amounts of training data, which is not always available for new tasks or domains. Several knowledge distillation methods have been proposed to address this issue that can train a smaller student network to mimic the prediction of a more extensive and more accurate teacher network.

Distillation has been applied in the field of intelligent systems, such as knowledge-based systems and rule-based systems, to reduce the system's size and improve the system's performance by improving the quality of the system's knowledge. The teacher and student models' difference can be considered a form of regularization, which is crucial to prevent overfitting. 

Algorithm \ref{alg:distillation} shows the pseudocode.

\begin{algorithm}
\caption{Pseudocode of distillation}
\label{alg:distillation}
\begin{algorithmic}
\STATE {\bfseries Input:} Dataset $D$, teacher model $T$, student model $S$, loss function $\mathcal{L}$, learning rate $\eta$, number of epochs $E$
\STATE {\bfseries Output:} Trained student model $S$
\STATE Initialize the weights of the student model $S$
\FOR{$e=1$ {\bfseries to} $E$}
\STATE Randomly shuffle the dataset $D$
\FOR{$i=1$ {\bfseries to} $|D|$}
\STATE Extract the $i^{th}$ sample $(x_i, y_i)$ from $D$
\STATE Forward propagate the sample $x_i$ through the teacher model $T$ to obtain the output probabilities $\hat{y}_i$
\STATE Compute the loss $\mathcal{L}$ using the output probabilities $\hat{y}_i$
\STATE Backpropagate the loss $\mathcal{L}$ through the student model $S$
\STATE Update the weights of the student model $S$ using the learning rate $\eta$
\ENDFOR
\ENDFOR
\STATE {\bfseries return} Trained student model $S$
\end{algorithmic}
\end{algorithm}

\subsection{Fundamental of Knowledge Distillation}

The fundamental of knowledge distillation is to transfer the knowledge from a large and deep neural network (teacher) to a small and shallow neural network (student). The knowledge distillation process consists of two steps: (1) training the teacher model and (2) distilling the knowledge from the teacher to the student (Figure~\ref{fig:knowledge_distillation_figure}).

\subsection{The Teacher Model}

The teacher model is typically a large and deep neural network, which is trained using a large dataset. The objective of the teacher model is to minimize the cross-entropy loss, which is defined as:
\begin{equation}
\mathcal{L}_{\mathrm{CE}} = -\frac{1}{N}\sum_{i=1}^{N}\sum_{j=1}^{C}y_{i,j}\log\hat{y}_{i,j},
\end{equation}
where $N$ is the number of samples, $C$ is the number of classes, $y_{i,j}$ is the ground truth label of the $i^{th}$ sample for the $j^{th}$ class, and $\hat{y}_{i,j}$ is the predicted probability of the $i^{th}$ sample for the $j^{th}$ class.

\subsection{Distilling the Knowledge}

The student model is typically a small and shallow neural network, which is trained using the teacher model's output probabilities, the teacher model's activations, or the intermediate representations of the teacher model. The distillation can be performed using a distillation loss, typically a combination of the cross-entropy loss and the distillation loss. The cross-entropy loss is used to minimize the difference between the output probabilities of the teacher and student models. In contrast, the distillation loss is used to reduce the difference between the intermediate representations of the teacher and student models.

The cross-entropy loss is defined as:
\begin{equation}
\mathcal{L}_{\mathrm{CE}} = -\frac{1}{N}\sum_{i=1}^{N}\sum_{j=1}^{C}y_{i,j}\log\hat{y}_{i,j},
\end{equation}
where $N$ is the number of samples, $C$ is the number of classes, $y_{i,j}$ is the ground truth label of the $i^{th}$ sample for the $j^{th}$ class, and $\hat{y}_{i,j}$ is the predicted probability of the $i^{th}$ sample for the $j^{th}$ class.

The distillation loss is typically defined as the Kullback-Leibler (KL) divergence between the output probabilities of the teacher and student models:
\begin{equation}
\mathcal{L}_{\mathrm{KL}} = \frac{1}{N}\sum_{i=1}^{N}\sum_{j=1}^{C}\hat{y}_{i,j}\log\frac{\hat{y}_{i,j}}{\tilde{y}_{i,j}},
\end{equation}
where $N$ is the number of samples, $C$ is the number of classes, $\hat{y}_{i,j}$ is the predicted probability of the $i^{th}$ sample for the $j^{th}$ class, and $\tilde{y}_{i,j}$ is the predicted probability of the $i^{th}$ sample for the $j^{th}$ class.

The distillation loss is typically defined as the mean squared error (MSE) between the activations of the teacher and student models:
\begin{equation}
\mathcal{L}_{\mathrm{MSE}} = \frac{1}{N}\sum_{i=1}^{N}\sum_{j=1}^{C}\left(\hat{a}_{i,j}-\tilde{a}_{i,j}\right)^{2},
\end{equation}
where $N$ is the number of samples, $C$ is the number of classes, $\hat{a}_{i,j}$ is the activation of the $i^{th}$ sample for the $j^{th}$ class, and $\tilde{a}_{i,j}$ is the activation of the $i^{th}$ sample for the $j^{th}$ class.

The distillation loss is typically defined as the MSE between the intermediate representations of the teacher and student models:
\begin{equation}
\mathcal{L}_{\mathrm{MSE}} = \frac{1}{N}\sum_{i=1}^{N}\sum_{j=1}^{C}\left(\hat{r}_{i,j}-\tilde{r}_{i,j}\right)^{2},
\end{equation}
where $N$ is the number of samples, $C$ is the number of classes, $\hat{r}_{i,j}$ is the intermediate representation of the $i^{th}$ sample for the $j^{th}$ class, and $\tilde{r}_{i,j}$ is the intermediate representation of the $i^{th}$ sample for the $j^{th}$ class.


Gradient-based untargeted attacks are significantly reduced using this technique. The employment of the standard goal function is no longer viable due to the effect of defense distillation on decreasing gradients to zero.

\section{Dataset Description and  Channel Estimation Scenario}\label{sec:dataset}
MATLAB 5G Toolbox provides a wide range of reference examples for next generation networks communications systems, such as 5G \cite{Matlab5G}. It also allows to customize and generate several types of waveforms, antennas and channel models to obtain datasets for DL-based models.  
In this study, the dataset used to train the DL-based channel estimation models is generated through a reference example in MATLAB 5G Toolbox, i.e, "Deep Learning Data Synthesis for 5G Channel Estimation". In the example, a convolutional neural network (CNN) is used for channel estimation. Single-input single-output (SISO) antenna method is also used by utilizing the physical downlink shared channel (PDSCH) and demodulation reference signal (DM-RS) to create the channel estimation model. 

The reference example in the toolbox generates 256 training datasets, i.e., transmit/receive the signal 256 times, for DL-based channel estimation model. Each dataset consists of 8568 data points, i.e., 612 subcarriers, 14 OFDM symbols,  1 antenna. However, each data point of training dataset is converted from a complex (real and imaginary) 612-14 matrix into a real-valued 612-14-2 matrix for providing inputs separately into the neural network during the training process. This is because the resource grids consisting of complex data points with real and imaginary parts in the channel estimation scenario, but CNN model manages the resource grids as 2-D images with real numbers. In this example, the training dataset is converted into 4-D arrays, i.e., 612-14-1-2N, where N presents the number of training examples, i.e., 256. \\
For each set of the training dataset, a new channel characteristic is generated based on various channel parameters, such as delay profiles (TDL-A, TDL-B, TDL-C, TDL-D, TDL-E), delay spreads (1-300 nanosecond), doppler shifts (5-400 Hz), and Signal-to-noise ratio (SNR or S/N) changes between 0 and 10 dB. Each transmitted waveform with the DM-RS symbols is stored in the train dataset, and the perfect channel values in train labels. The CNN-based channel estimation based is trained with the generated dataset. MATLAB 5G toolbox also allows tuning several communication channel parameters, such as the frequency, subcarrier spacing, number of subcarriers, cyclic prefix type, antennas, channel paths, bandwidth,  code rate, modulation, etc. The channel estimation scenario parameters with values are given for each  in Table  \ref{tab:dataset}.

\begin{table}[!htbp]
\centering
\caption{The channel estimation parameters with values}
\label{tab:datasetParameter}
\begin{tabular}{|l|c|}
\hline
\textbf{Channel Parameter}& {\textbf{Value }}  \\
 \hline 
Delay Profile & TDL-A, TDL-B, TDL-C, TDL-D, TDL-E \\ \hline
Delay Spread & 1-300 ns \\ \hline
Maximum Doppler Shift & 5-400 Hz\\  \hline
NFFT& 1024  \\ \hline
Sample Rate& 30720000  \\  \hline
Symbols Per Slot& 14 \\  \hline
Windowing& 36 \\  \hline
Slots Per Subframe& 2 \\ \hline
Slots Per Frame& 20 \\ \hline
Polarization & Co-Polar \\ \hline
TransmissionDirection & Downlink \\ \hline
NumTransmitAntennas & 1 \\ \hline
NumReceiveAntennas & 1 \\ \hline
FadingDistribution & Rayleigh\\ \hline
Modulation & 16QAM\\ \hline
\end{tabular}
\label{tab:dataset}
\end{table}

The training dataset is split into validation and training sets due to avoiding to overfit the training data. Training set is used to train and fit the model, while the validation data is used for monitoring the performance of the trained neural network at  certain intervals, i.e., 5 per epoch. It is expected that the training stop when the validation loss stops decreasing and improving the model. In this study, most part of dataset is used for training, i.e., 80\% for training, and 20\% for
testing.
  
\section{Performance Metric}
The performance metric, MSE (Mean Squared Error), is used to evaluate and compare CNN-based models. The
MSE scores are utilized for further analyses of the model. MSE equation is given below. It measures the average squared difference between the actual and predicted values. The MSE equals zero when a model has no error. Model error increases along with  the MSE value.

 \begin{equation}
\begin{split}
MSE = {\frac{\sum_{}^{}{(Y_{t} - {\hat{Y}}_{t})}^{2}}{n}}
\end{split}
\end{equation}

Where : \(Y_{t}\) :The actual t\textsuperscript{th} instance, \({\hat{Y}}_{t}\ \): The forecasted t\textsuperscript{th} instance, n: The total number of instance

\section{System Overview}
\label{sec:overview}

\subsection{Complex Numbers and Wireless Communication}
In mathematics, a complex number is a number that can be expressed in the form $a + bi$, where $a$ and $b$ are real numbers and $i$ is the imaginary unit, a number such that $i^2 = -1$. Complex numbers are used in a wide range of fields, including mathematics, physics, engineering, and economics. The imaginary unit $i$ is used to denote a square root of $-1$. 
Complex numbers are used in wireless communication technologies. The complex number system modifies and demodulates wireless signals in digital wireless communication. The most significant distinction between the real and complex number systems is that the complex number system contains more than one dimension. Adversarial ML attacks, on the other hand, use real numbers to enter the decision boundaries of the victim DL models, and the final malicious inputs are in the real number domain. To solve this challenge, complex numbers are split into their real and imaginary elements. Table \ref{tab:conversion} shows the example dataset.

\begin{table}[!htbp]
    \centering
        \caption{Example dataset. The original dataset is shown as complex numbers in the table at the top. The training dataset is represented in real numbers in the table below.}
    \begin{tabular}{|c|c|c|c|}
\hline
 \textbf{F1} & \textbf{F2} & \textbf{F3} & \textbf{F4}\\
\hline \hline
 0.15+0.90j &  0.26+0.90j &  0.32+0.90j &  0.41+0.88j \\
-0.39-0.84j & -0.46-0.83j & -0.55-0.79j & -0.61-0.72j \\
-0.26-0.89j & -0.38-0.87j & -0.44-0.84j & -0.50-0.80j \\
-0.56+0.78j & -0.45+0.82j & -0.37+0.89j & -0.28+0.89j \\
$\vdots$ & $\vdots$ & $\vdots$ & $\vdots$ \\
-0.86-0.43j & -0.88-0.35j & -0.87-0.23j & -0.89-0.12j \\
\hline
\end{tabular}

\vspace{10pt}
{\Large $\Downarrow$}
\vspace{10pt}

    \begin{tabular}{|c|c||c|c||c|c||c|c|}
\hline
 \textbf{F1-1} & \textbf{F1-2} & \textbf{F2-1} & \textbf{F2-2} & \textbf{F3-1} & \textbf{F3-2} & \textbf{F4-1} & \textbf{F4-2}\\
\hline \hline
 0.15 & 0.90 & 0.26 & 0.90  &  0.32 & 0.90 &  0.41 & 0.88 \\
-0.39 & 0.84 & -0.46 & 0.83 & -0.55 & 0.79 & -0.61 & 0.72 \\
-0.26 & 0.89 & -0.38 & 0.87 & -0.44 & 0.84 & -0.50 & 0.80 \\
-0.56 & 0.78 & -0.45 & 0.82 & -0.37 & 0.89 & -0.28 & 0.89 \\
$\vdots$ & $\vdots$ & $\vdots$ & $\vdots$ & $\vdots$ & $\vdots$ & $\vdots$ & $\vdots$  \\
-0.86 & 0.43 & -0.88 & 0.35 & -0.87 & 0.23 & -0.89 & 0.12 \\
\hline
\end{tabular}
    \label{tab:conversion}
\end{table}

\subsection{System Model}

Figure \ref{fig:cnn_model_overview} shows the CNN based DL model used in this paper for the channel estimation. The input to the model is the pilot signals with different subcarriers and OFDM symbols. The input is first passed through a convolutional layer, followed by a max-pooling layer. The output of the max-pooling layer is then passed through a fully connected layer, followed by a softmax layer. The final output of the model is the channel estimation.

\section{Evaluation and Performance Results}
\label{sec:experiments}

\textcolor{black}{This section provides performance results} to evaluate the proposed mitigation methods for DL-based channel estimation models in next-generation networks. We use the channel estimation dataset described in Section \ref{sec:dataset} to train the model. We use five different attacks (i.e., FGSM, BIM, MIM, PGD and C\&W) to evaluate the proposed mitigation methods. The deep learning-based channel estimation model is trained in the TensorFlow environment. The proposed mitigation methods are implemented in the Keras environment. The MSE performance metric is used to evaluate the accuracy of the channel estimation model.

\subsection{Experimental Setup}

The teacher and student models are a DNN with 3 convolutional layers. The teacher and student models are trained using stochastic gradient descent with a momentum of 0.9 and a learning rate of 0.001. The teacher and student models are trained for 100 epochs. The batch size is set to 256. Table \ref{tab:cnn_params} shows the DL model hyper-parameters.

\begin{table}[!htbp]
    \caption{CNN Model architecture parameters for the teacher and the student models}
    \label{tab:cnn_params}
    \centering
    \begin{tabular}{|c|c|c|c|p{1cm}|c|c|c|}
    \hline
          & \textbf{Name} & \textbf{Type} & \textbf{Filters} & \textbf{Kernel Size} & \textbf{Padding} & \textbf{Output}  \\ \hline\hline
         \multirow{3}{*}{\rotatebox[]{90}{\textbf{Teacher}}}& Conv-1 & Conv2D & 48 & (9, 9) & same & (1, 612, 48) \\
         & Conv-2 & Conv2D & 16 & (5, 5) & same & (1, 612, 16) \\
         & Conv-3 & Conv2D & 1 & (5, 5) & same & (1, 612, 1) \\ \hline \hline
         \multirow{3}{*}{\rotatebox[]{90}{\textbf{Student}}} & Conv-1 & Conv2D & 24 & (9, 9) & same & (1, 612, 48) \\
         & Conv-2 & Conv2D & 8 & (5, 5) & same & (1, 612, 16) \\
         & Conv-3 & Conv2D & 1 & (5, 5) & same & (1, 612, 1) \\
         \hline
    \end{tabular}
\end{table}

\begin{figure*}[!htbp]
    \centering
    \includegraphics[width=0.7\linewidth]{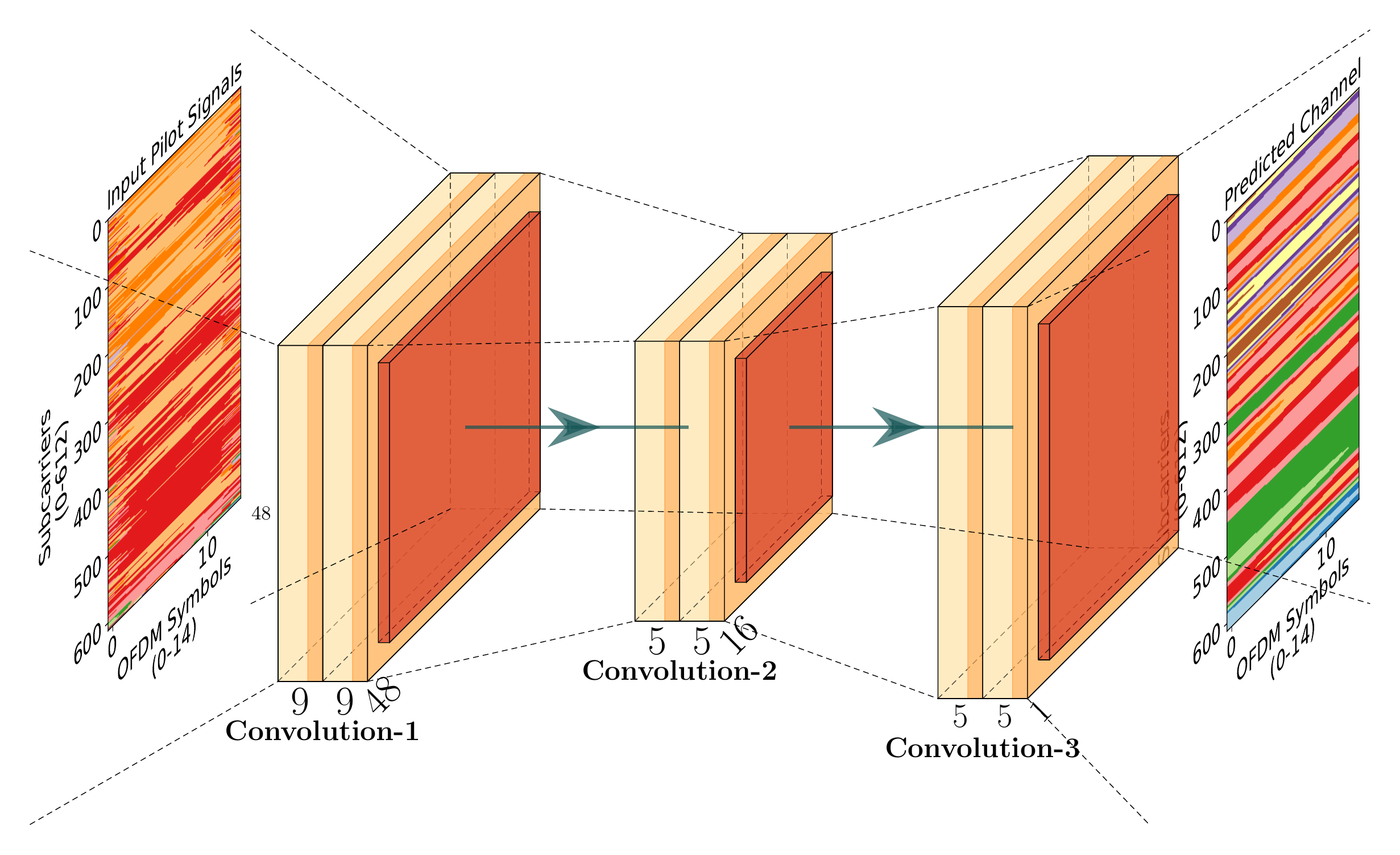}
    \caption{The vulnerable CNN model based channel estimation overview}
    \label{fig:cnn_model_overview}
\end{figure*}

Figure \ref{fig:cnn_model_overview} shows the architecture of the teacher and student models.

\textcolor{black}{The models are generative and supervised models trained to predict channel parameters defined at the receiver. The input and output size is $612 \times 14$ (e.g. $Subcarriers \times OFDM \, symbols$). Table \ref{tab:cnn_hyperparams} shows the CNN model's hyper-parameters.}

\begin{table}[!htbp]
    \centering
    \caption{\textcolor{black}{CNN Model architecture parameters for the teacher and the student models}}
    \label{tab:cnn_hyperparams}
    \begin{tabular}{|l|m{4cm}|}
    \hline
        \textbf{Hyper parameter} & \textbf{Value} \\ \hline \hline
        \textbf{Framing Problem} & Supervised Regression \\ \hline
        \textbf{Initialization method} & GlorotUniform \\ \hline
        \textbf{Activation functions} & \begin{itemize}
\item \textbf{Conv-1}: Selu
\item \textbf{Conv-2}: Softplus
\item \textbf{Conv-3}: Selu
\end{itemize} \\ \hline
        \textbf{Number of parameters} & \begin{itemize}
\item \textbf{Undefended}: 6977
\item \textbf{Teacher}: 23533
\item \textbf{Student}: 6977
\end{itemize}\\
         \hline
    \end{tabular}
    
\end{table}

\subsection{Performance Results}
This section provides the results for the proposed defensive distillation-based mitigation method. We applied the attack success ratio (ASR) as the performance metric. ASR is the ratio of test samples that an attacker can mispredict to the total number of test samples. The highest ASR indicates that the attack is more effective. The following equation is used to calculate ASR:

\begin{equation}
\text{ASR} = \frac{1}{m}\sum_{i=0}^m{\frac{MSE(\mathbf{x}^{adv}_{(i)},\mathbf{y}_{(i)})-MSE(\mathbf{x}_{(i)},\mathbf{y}_{(i)})}{MSE(\mathbf{x}^{adv}_{(i)},\mathbf{y}_{(i)})}}
\end{equation}

\textcolor{black}{
Table \ref{tab:init_mse_vals} shows the initial prediction performance results of all models with test dataset.}

\begin{table}[!htbp]
    \centering
    \caption{\textcolor{black}{Initial MSE values with test (i.e. benign) dataset}}
    \label{tab:init_mse_vals}
    \begin{tabular}{|c|c|} \hline 
        \textbf{Model} &  \textbf{MSE}\\ \hline \hline 
        Undefended & 0.02766 \\
        Teacher & 0.02484 \\
        Student & 0.02558 \\ \hline 
    \end{tabular}
\end{table}

The first experiment is to perform attacks on the undefended model, as shown in Table \ref{tab:undef_model_res}.

\begin{table}[!htbp]
    \centering
    \caption{Experimental results for the undefended DL model. The results show that the initial DL model is vulnerable to the adversarial ML attacks.}
    \label{tab:undef_model_res}
    \begin{tabular}{|l|l|r|r|c|}
\hline
\multicolumn{2}{|c|}{} & \multicolumn{2}{|c|}{\textbf{MSE}} & \multirow{2}{*}{\textbf{ASR}}\\ \cline{1-4} 
\textbf{Attack} & $\epsilon$ &  \textbf{Benign Input}  & \textbf{Malicious Input}  &   \\
\hline \hline
\multirow{5}{*}{\textbf{BIM}} & 0.1 &            0.028126 &                 0.028485 &              0.018932 \\
    & 0.5 &            0.028128 &                 0.036766 &              0.289385 \\
    & 1.0 &            0.028106 &                 0.073039 &              0.613742 \\
    & 2.0 &            0.028222 &                 0.192034 &              0.832142 \\
    & 3.0 &            0.027837 &                 0.306523 &              0.904284 \\ \hline \hline
\multirow{5}{*}{\textbf{FGSM}} & 0.1 &            0.028213 &                 0.028477 &              0.013840 \\
    & 0.5 &            0.028223 &                 0.034714 &              0.215770 \\
    & 1.0 &            0.028106 &                 0.052979 &              0.433404 \\
    & 2.0 &            0.028121 &                 0.121161 &              0.617207 \\
    & 3.0 &            0.028126 &                 0.234474 &              0.689940 \\ \hline \hline
\multirow{5}{*}{\textbf{MIM}} & 0.1 &            0.028126 &                 0.028493 &              0.019298 \\
    & 0.5 &            0.028229 &                 0.037301 &              0.297863 \\
    & 1.0 &            0.027990 &                 0.069112 &              0.599503 \\
    & 2.0 &            0.028228 &                 0.162054 &              0.825845 \\
    & 3.0 &            0.028228 &                 0.323735 &              0.908491 \\ \hline \hline
\multirow{5}{*}{\textbf{PGD}} & 0.1 &            0.028000 &                 0.028363 &              0.019231 \\
    & 0.5 &            0.028205 &                 0.036839 &              0.288993 \\
    & 1.0 &            0.028106 &                 0.073028 &              0.613850 \\
    & 2.0 &            0.028141 &                 0.192549 &              0.833080 \\
    & 3.0 &            0.027913 &                 0.317048 &              0.905127 \\
\hline \hline
\textbf{C\&W} & - & 0.028314 & 0.029803 & 0.066435 \\ \hline
\end{tabular}
    
\end{table}

The results of the first experiment show that the initial DL model is vulnerable to the adversarial ML attacks. As expected, the ASR value has a positive correlation with $\epsilon$ value. The results also show that the BIM, MIM, and PGD attacks are more effective than the FGSM and C\&W (without $\epsilon$) attacks model under the same $\epsilon$. The success rate of the C\&W attack model is lower than that of the BIM, MIM, and PGD attack models. 

Experimental results for the proposed defensive distillation-based
mitigation method are shown in Table \ref{tab:distilled_res}. 

\begin{table}[!htbp]
    \centering
    \caption{Experimental results for the proposed defensive distillation-based mitigation method. The results show that the proposed method can improve the accuracy of the channel estimation model. The results indicate that the proposed method can provide better results for the attacks (i.e., FGSM, BIM, MIM, PGD, and C\&W).}
    \label{tab:distilled_res}
    \begin{tabular}{|l|l|r|r|c|}
\hline
\multicolumn{2}{|c|}{} & \multicolumn{2}{|c|}{\textbf{MSE}} & \multirow{2}{*}{\textbf{ASR}}\\ \cline{1-4} 
\textbf{Attack} & $\epsilon$ &  \textbf{Benign Input}  & \textbf{Malicious Input}  &   \\
\hline \hline
\multirow{5}{*}{\textbf{BIM}} & 0.1 &            0.027861 &                 0.028192 &              0.018048 \\
    & 0.5 &            0.027859 &                 0.029179 &              0.066479 \\
    & 1.0 &            0.027857 &                 0.029177 &              0.066474 \\
    & 2.0 &            0.027860 &                 0.029179 &              0.066478 \\
    & 3.0 &            0.027865 &                 0.029185 &              0.066460 \\ \hline \hline
\multirow{5}{*}{\textbf{FGSM}} & 0.1 &            0.027851 &                 0.027854 &              0.000118 \\
    & 0.5 &            0.027853 &                 0.027921 &              0.003289 \\
    & 1.0 &            0.027845 &                 0.028108 &              0.012865 \\
    & 2.0 &            0.027851 &                 0.028870 &              0.047475 \\
    & 3.0 &            0.027851 &                 0.030105 &              0.095989 \\ \hline \hline
\multirow{5}{*}{\textbf{MIM}} & 0.1 &            0.027864 &                 0.028198 &              0.018295 \\
    & 0.5 &            0.027863 &                 0.029232 &              0.068893 \\
    & 1.0 &            0.027863 &                 0.029232 &              0.068896 \\
    & 2.0 &            0.027860 &                 0.029229 &              0.068908 \\
    & 3.0 &            0.027860 &                 0.029229 &              0.068914 \\ \hline \hline
\multirow{5}{*}{\textbf{PGD}} & 0.1 &            0.027859 &                 0.028190 &              0.018059 \\
    & 0.5 &            0.027866 &                 0.029183 &              0.066392 \\
    & 1.0 &            0.027865 &                 0.029183 &              0.066400 \\
    & 2.0 &            0.027857 &                 0.029175 &              0.066423 \\
    & 3.0 &            0.027862 &                 0.029180 &              0.066412 \\
\hline \hline
\textbf{C\&W} & - & 0.027263 & 0.027408 & 0.00793 \\ \hline
\end{tabular}
\end{table}
The experimental results show that the proposed method can improve the accuracy of the channel estimation model. The results also show that the proposed method can provide better results for the attacks (i.e., FGSM, BIM, MIM, PGD, and C\&W). 

\begin{figure*}[!htbp]
    \centering
    \includegraphics[width=0.7\linewidth]{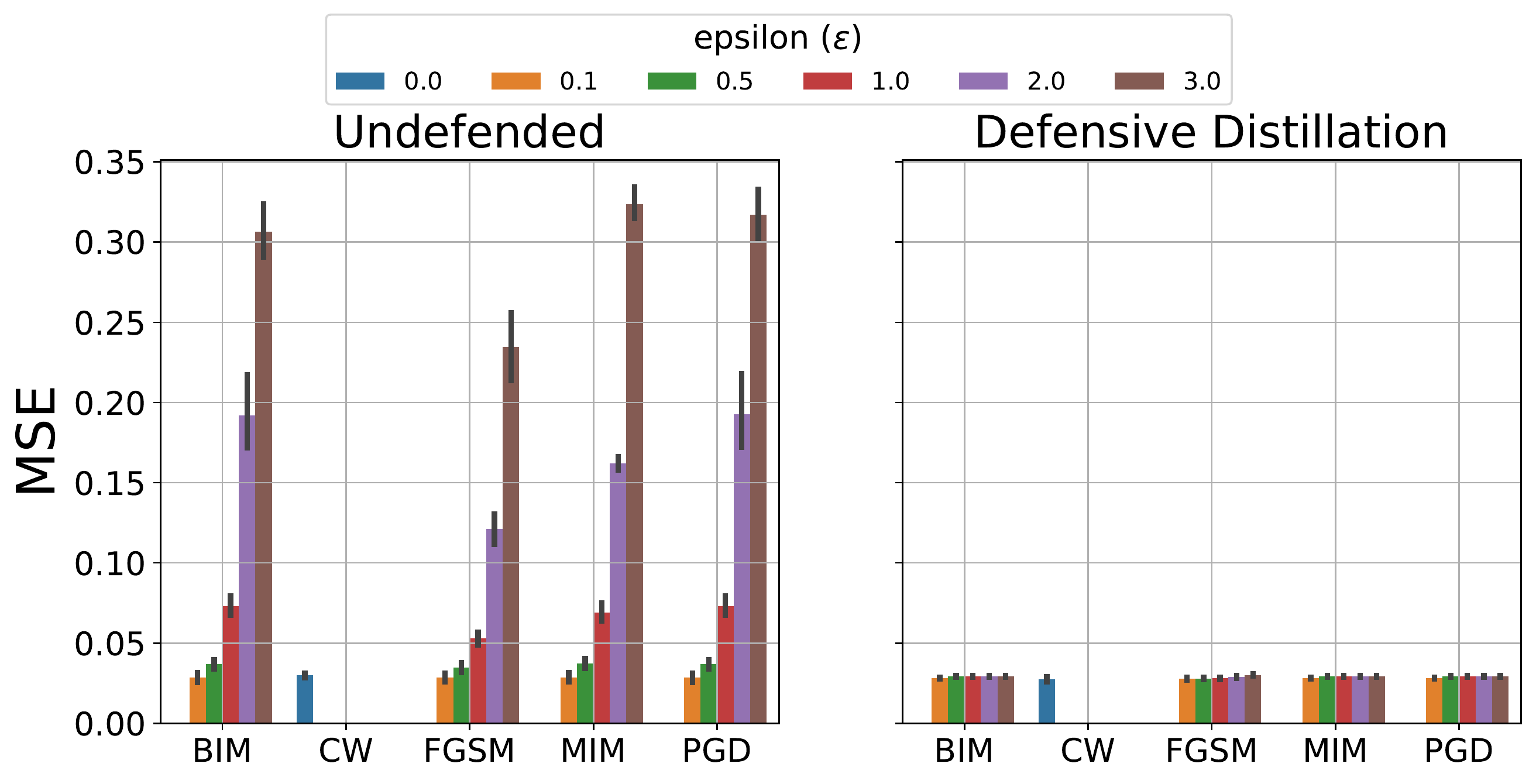}
    \caption{Experimental results for the proposed defensive distillation-based mitigation method. The results show that the proposed method can improve the accuracy of the channel estimation model}
    \label{fig:def_distilled_model}
\end{figure*}

Figure  \ref{fig:def_distilled_model} shows the MSE results with 6 different $\epsilon$ values (i.e. $0.0, 0.1, 0.5, 1.0, 2.0, 3.0$) for the undefended and defensive distillation based defended DL model for the  all attacks. Because there is no $\epsilon$ value for the C\&W attack, there is only one bar chart for the C\&W attack.  The results show that the proposed method can improve the accuracy of the channel estimation model.

\begin{figure*}[!htbp]
    \centering
    \includegraphics[width=1.0\linewidth]{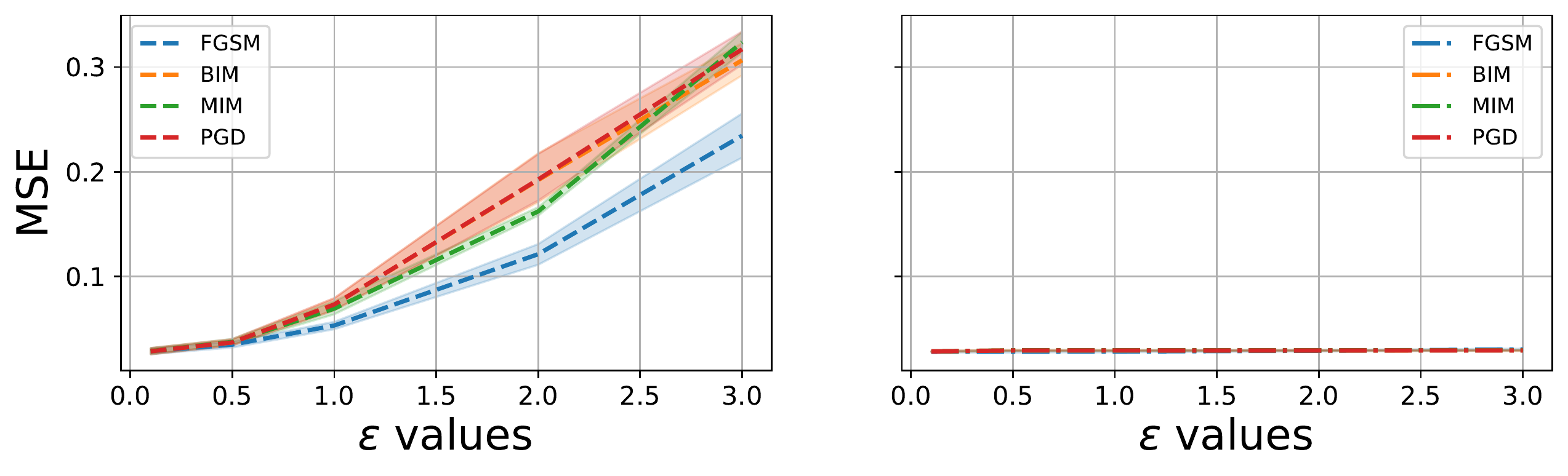}
    \caption{MSE trend line for the undefended and defended DL models. }
    \label{fig:def_distilled_model-2}
\end{figure*}

Figure \ref{fig:def_distilled_model-2} shows the MSE change with different $\epsilon$ values for each attack for the undefended and defensive distillation based defended DL model. The defended model's MSE values (i.e. the right figure) are almost similar to each attack and $\epsilon$ values. We can see that the defensive distillation based mitigation method works pretty well against all types of adversarial attacks.

\section{Discussion}
\label{sec:discussion}
This study provides a comprehensive analysis of the DL-based channel estimation model in terms of vulnerabilities. The model's vulnerabilities are studied for various adversarial attacks, incluing FGSM, BIM, PGD, MIM, and C\&W, as well as the mitigation method, i.e., defensive distillation.  The results show that CNN based channel estimation models are vulnerable against adversarial attacks, i.e., FGSM, BIM,
MIM, PGD, and C\&W. The attack success ratio is also pretty much high, i.e., 0.9, under a higher power attack ($\epsilon$ equals 3.0) for BIM, MIM, and PGD attacks. On the other hand, the rate is very low for C\&W attack, i.e., 0.06, comparing with the others. Fortunately,  the proposed defensive distillation-based mitigation method provides a better performance against higher-order adversarial attacks, and the attack success ratio goes down to 0.06 for BIM, MIM, and PGD attacks. The impact of mitigation method on FGSM is  lower than others, i.e., the attack succes rate is 0.09. For C\&W, the attack success rate goes from 0.06 to 0.007 after applying the proposed defensive distillation-based mitigation method. According to results, adversarial attacks on DL-based channel estimation models and the use of the proposed defensive distillation-based mitigation method can be summarized as:\\
\emph{\textit{Observation 1}}: The DL-based channel estimation models are vulnerable to adversarial attacks, especially BIM, MIM, and PGD.\\
\emph{\textit{Observation 2}}: BIM, MIM, and PGD attacks are the most successful
attacks in terms of the attack success rate.\\
\emph{\textit{Observation 3}}: The DL-based channel estimation models are more robust against C\&W attack.\\
\emph{\textit{Observation 4}}: A strong negative correlation exists between attack power  \(\epsilon\) and the performance of channel estimation models.\\
\emph{\textit{Observation 5}}: The proposed mitigation method, i.e., defensive distillation, offers a better performance against adversarial attacks.\\

\section{Conclusion and Future Works}
\label{sec:conclusion}
The mobile wireless communication networks are rapidly developing with the high demand and advanced communication and computing technologies. The last few years have experienced a remarkable growth in wireless industry, especially for NextG networks.  This paper provides a comprehensive vulnerability analysis of deep learning (DL) based channel estimation models for adversarial
attacks (i.e., FGSM, BIM, PGD, MIM, and C\&W) and defensive distillation
based mitigation method in NextG networks.  The results confirm that the original DL-based channel estimation model is significantly vulnerable to adversarial attacks, especially BIM, MIM, and
PGD. The attack success rare increases under a heavy adversarial attack (\(\epsilon\)= 3.0) up to 0.9 for those attacks. There is a high positive correlation between attack power  \(\epsilon\) and the attack success rate as expected, i.e., a high \(\epsilon\) increases as the the attack success rate. On the other hand,the proposed defensive distillation-based mitigation method can improve the accuracy of the channel estimation model, and provide better results against higher-order adversarial attacks, e.g., the attack success rate goes from 0.9 to 0.06 after applying the proposed mitigation method. The overall results prove that the
proposed method can provide better results for the attacks (i.e., FGSM, BIM, MIM, PGD, and C\&W) in terms of the model accuracy and the attack success rate.

As future work, the authors will focus on the Intelligent Reflecting Surface (IRS) and spectrum sensing using AI-based models, and their cybersecurity risks.

%

\bibliography{refs}

\bibliographystyle{IEEEtran}

\begin{IEEEbiography}[{\includegraphics[width=1in,height=1.25in,clip,keepaspectratio]{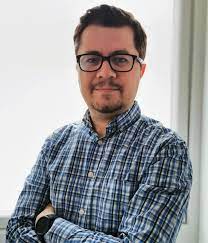}}]{Ferhat Ozgur Catak} is an associate professor at the University of Stavanger, Norway. He completed his B.Sc.  degree in electrical/electronic engineering in 2002 and his Ph.D. degree in Informatics in 2014. Previously, he worked at TUBITAK in Turkey, NTNU, and Simula Research laboratory in Norway. His research areas are cyber security, malware analysis, secure multi-party computation, and privacy methods.
\end{IEEEbiography}

\begin{IEEEbiography}[{\includegraphics[width=1in,height=1.25in,clip,keepaspectratio]{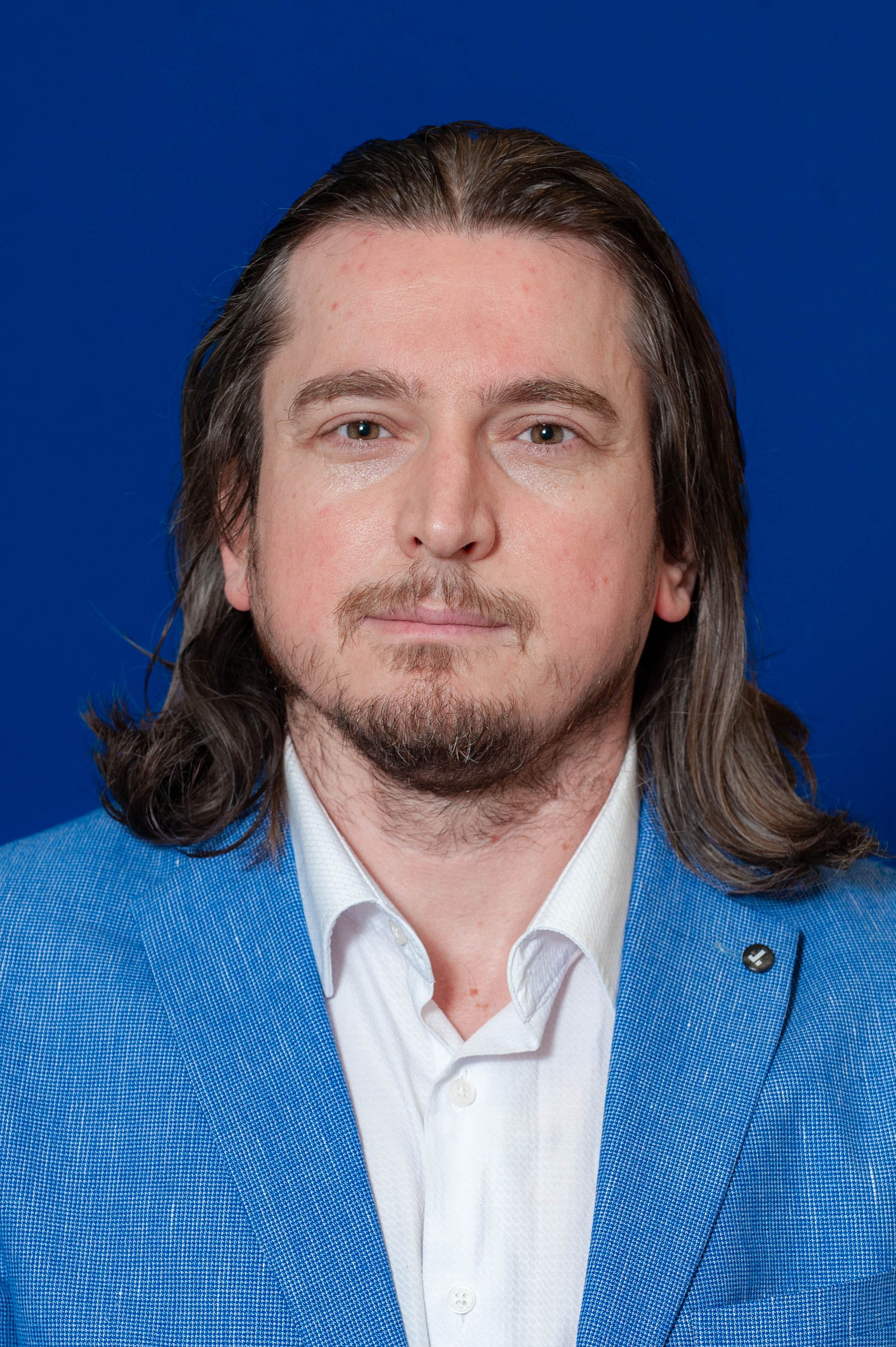}}]{Murat Kuzlu} (Senior Member, IEEE) joined the Department of Engineering Technology,  Old Dominion University (ODU)  in 2018 as an Assistant Professor. He received his B.Sc., M.Sc., and Ph.D. degrees in Electronics and Telecommunications Engineering from Kocaeli University, Turkey, in 2001, 2004, and 2010, respectively. From 2006 to 2011, he worked as a senior researcher at TUBITAK-MAM (Scientific and Technological Research Council of Turkey – The Marmara Research Center). Before joining ODU, he worked as a Research Assistant Professor at Virginia Tech’s Advanced Research Institute from 2011 to 2018. His research interests include cyber-physical systems, smart cities, smart grids, artificial intelligence, and next generation wireless networks. 
\end{IEEEbiography}

\begin{IEEEbiography}[{\includegraphics[width=1in,height=1.25in,clip,keepaspectratio]{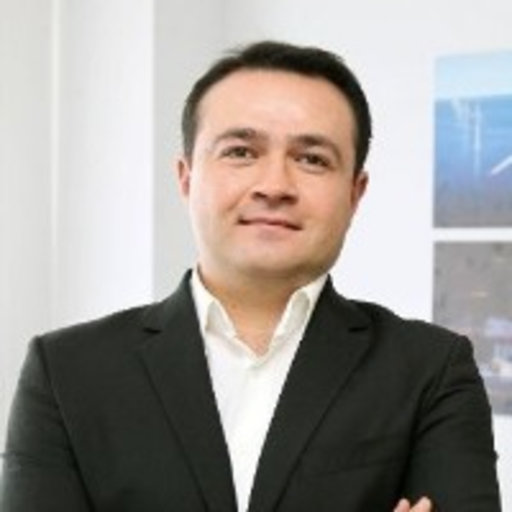}}]{Umit Cali} joined the Department of Electric Power Engineering, Norwegian University of Science and Technology, Norway in 2020 as an associate professor. He received the B.E. degree in electrical engineering from Yildiz Technical University, Istanbul, Turkey, in 2000, and the M.Sc. degree in electrical communication engineering and the Ph.D. degree in electrical engineering and computer science from the University of Kassel, Germany, in 2005 and 2010, respectively. He worked at the University of Wisconsin-Platteville and the University of North Carolina at Charlotte as an Assistant Professor between 2013 and 2020, respectively. His current research interests include energy informatics, artificial intelligence, blockchain technology, renewable energy systems, and energy economics. He is serving as an active Vice-Chair of the IEEE Blockchain in Energy Standards WG (P2418.5).
\end{IEEEbiography}

\begin{IEEEbiography}[{\includegraphics[width=1in,height=1.25in,clip,keepaspectratio]{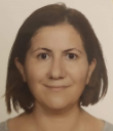}}]{Evren Catak} received the B.Sc. degree in Electrical and Electronics Engineering from Eskisehir Osmangazi University, Turkey in 2002, the M.Sc. degree in Electronics Engineering from Kadir Has University, Istanbul, Turkey in 2012, and the Ph.D. degree in Communication Engineering from Yildiz Technical University, Istanbul, Turkey in 2017. She was a postdoctoral fellow at the Norwegian University of Science and Technology. Her research interests are in the physical layer design of emerging communication systems, communication theory, signal processing, and wireless communications
\end{IEEEbiography}

\begin{IEEEbiography}[{\includegraphics[width=1in,height=1.25in,clip,keepaspectratio]{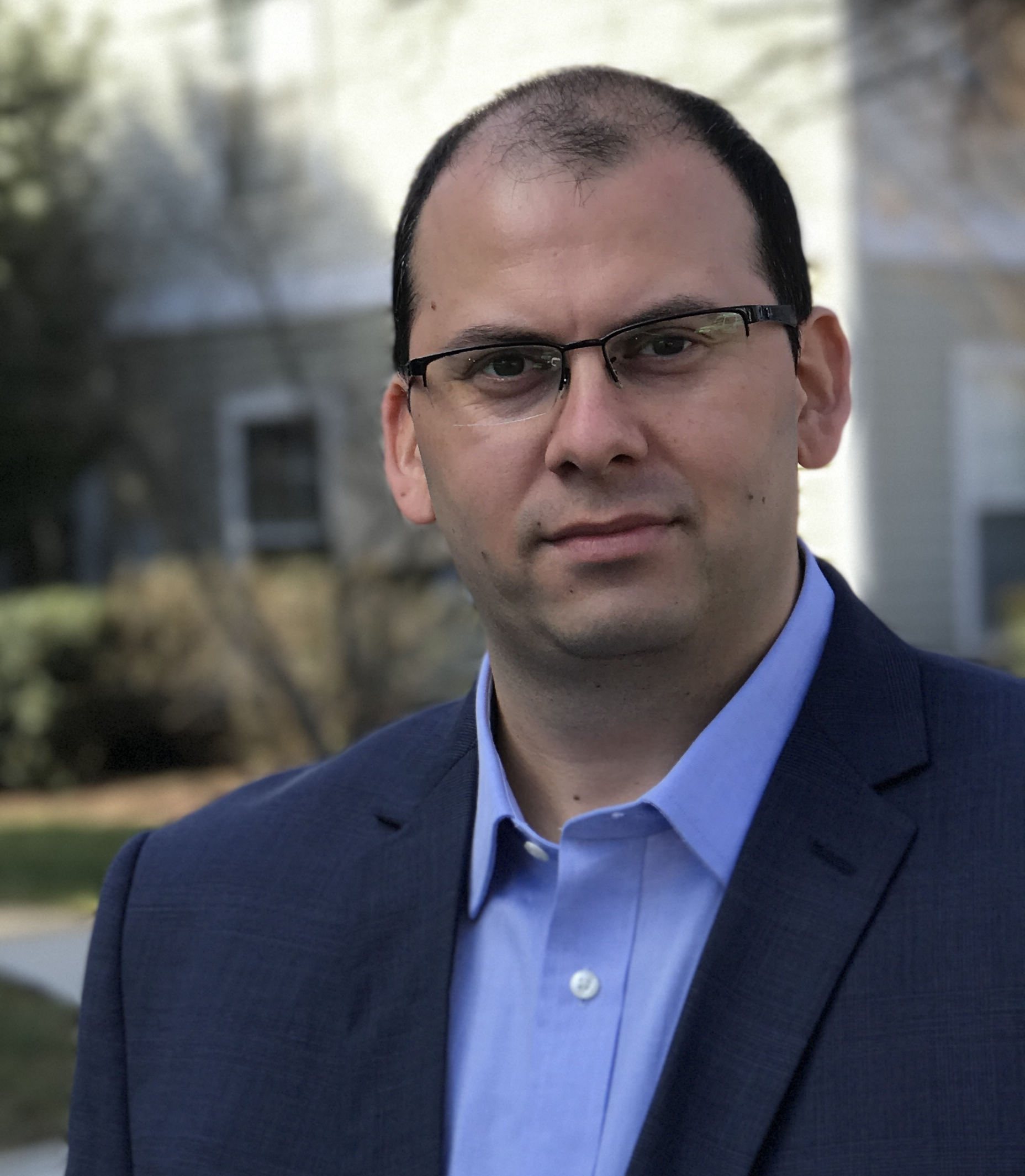}}]{Ozgur Guler} received the B.S. degree in computer science and the M.S. degree in computer science with a focus on image-guided surgery from the University of Innsbruck, Innsbruck, Austria, and the Ph.D. degree from the Medical University of Innsbruck, Austria, with a focus on image-guided diagnosis and therapy. He is currently an Imaging Scientist and AI Researcher specialized in 3-D chronic wound imaging and computer vision. Prior to joining eKare Inc., he was a Researcher with the Sheikh Zayed Institute (SZI) for Pediatric Surgical Innovation Center, Washington, DC, where he developed the segmentation and classification algorithms that laid the groundwork of the eKare inSight system.
\end{IEEEbiography}

\end{document}